# Geometric measure for two-qubit CS-states: an exactly solvable model

ABOLFATHIYAR MASOUD[1,*], AND ENTEKHABI SINA[2,*#]

[1]*Departement of Physics, Faculty of Sciences, Azerbaijan University of Shahid Madani, Tabriz, Iran, e.mail:*
*mphysic1365@gmail.com*

[2]*Department of Computer Sciences, Faculty of Engineering, Urmia University, Urmia, Iran, e.mail:*
*cinarooney@yahoo.com*

## ABSTRACT

Newly, during an analytical calculation M. A. Yurischev [arXiv: 1302.5239 (2013)] prove that quantum discord (or entanglement) remains unchangeable for any centrosymmetric (CS) states that are converted to their corresponding mode X. In this paper, in addition to the above calculations, the conditions which geometric measure remains constant are presented. This work will be completed with two physical examples.

*Key word*: density matrix; geometric measure; centrosymmetric; correlation.

## INTRODUCTION

As we know, correlation can be measured by several different measures which in its turn exhibit certain behaviors, but due to problems are seen in computing and behavioral concepts, none of them can be made publicly. Among these measures the entanglement (Yu, 2004; Yu, 2006; Yu, 2007; Be3llomo *et al.,* 2007; Luo, *et al.*, 2011; Ma *et al.,* 2012; Yönac *et al., 2007*), the negativity (Silva *et al.,* 2013*;* Zhang *et al.,* 2010)), quantum discord (Ollivier & Zurek, 2001; Luo, 2008; Ali *et al.,* 2010; Dakic *et al.,* 2010; Luo & Fu, 2010), and also geometric measure of quantum discord (Shi *et al.,* 2011) can be noted.

In some physical problems to get the density matrices to calculate their quantum correlation, we will encounter a lot of problems. So, in a way computations that are done on the matrix, can be done on another matrix too, that is obtained by the local orthogonal transformation conversion, while the new matrix plays the same role in previous matrix. It would become the CS-state to the X-state and vice versa as follows (Yurischev, 2013):

$$H \otimes H \rho_{CS} H \otimes H = \rho_X$$
$$H \otimes H \rho_X H \otimes H = \rho_{CS}$$

where $H = \frac{1}{\sqrt{2}}\begin{pmatrix} 1 & 1 \\ 1 & -1 \end{pmatrix}$ is the Hadamard transformation.

With an analytical solution that has been shown quantum discord behavior is identical to the CS and X modes. We will prove here that geometric measure (G) taking into account the specific conditions can be treated the same for the X and CS modes. Meantime, we will try to present calculates in cases where it seems suspicious to ensure that the above conditions are provided. Finally, two physical examples will be given to show the practical concepts of conditions which will be mentioned.

# MAIN RESULTS

## Conclusion

In a nutshell, maybe in any problems we deal with CS-type density matrices, as respects computing of the quantum correlation for these type matrices are very difficult and sometimes impossible, therefore, it is better that CS-type matrix become X-type (Yurischev, 2013), because calculations of X-type matrix are certain. Such calculations remain unchangeable during the conversion of quantum discord, but in this work, we proved that geometric measure also remains unchangeable under some conditions (6), while both of them are the quantum correlation measures. But probably computations of G have never been easier than before, and we only show that if for instance computing of G deal with problem in X-states we can calculate those in CS-states and vice versa. Briefly, we obtain $\lambda_M^{\max}$, and finally we have shown that geometric measure could be calculated for them by two physical examples (Yurischev, 2013; Chen & Zhi, 1998; Fel'dman, 2012).

# APPLICATIONS

## The same geometric measure

A general CS density matrix $4 \times 4$ with seven real parameters is written as follows (Yurischev, 2013):

$$\rho_{CS} = \begin{pmatrix} p_1 & p_2 + ip_3 & p_4 + ip_5 & p_6 \\ p_2 - ip_3 & \frac{1}{2} - p_1 & p_7 & p_4 - ip_5 \\ p_4 - ip_5 & p_7 & \frac{1}{2} - p_1 & p_2 - ip_3 \\ p_6 & p_4 + ip_5 & p_2 + ip_3 & p_1 \end{pmatrix} \quad (1)$$

And also the most general X density matrix with seven real parameters can be written as follows:

$$\rho_X = \begin{pmatrix} q_1 & 0 & 0 & q_4 + iq_5 \\ 0 & q_2 & q_6 + iq_7 & 0 \\ 0 & q_6 - iq_7 & q_3 & 0 \\ q_4 - iq_5 & 0 & 0 & 1 - q_1 - q_2 - q_3 \end{pmatrix} \quad (2)$$

Definition of G is based on a " distance" that it can be calculated in different ways, such as trace distance, Hilbert-Schmidt distance, Bures distance and else. But the distance that we will use here, is HS distance (or the abbreviation HS). In accordance with the method that we shall use [13], first the matrix R with the elements $R_{\mu\nu} = Tr\left[\rho(\sigma_\mu \otimes \sigma_\nu)\right]; \mu,\nu = 0,1,2,3$, must be written for each of the modes (where $\sigma_0$ is $2 \times 2$ identity matrix and $\sigma_i$; i=1,2,3 are Pauli matrices), that (3) and (4) are related to them. Then, by using local Neumann measurements for each of them, measurement-induced classical-classical (MICC) obtain. Finally the correlation obtains by using HS between the initial state and MICC state (Shi et al., 2011).

$$R_{\rho_{CS}} = \begin{pmatrix} 1 & 4p_2 & 0 & 0 \\ 4p_4 & 2(p_6+p_7) & 0 & 0 \\ 0 & 0 & 2(p_6-p_7) & -4p_5 \\ 0 & 0 & -4p_3 & 4p_1-1 \end{pmatrix} \quad (3)$$

$$R_{\rho_X} = \begin{pmatrix} 1 & 0 & 0 & 2(q_1+q_3)-1 \\ 0 & 2(q_6+q_4) & 0 & 0 \\ 0 & 0 & 2(q_6-q_4) & 0 \\ 2(q_1+q_2)-1 & 0 & 0 & 1-2(q_2+q_3) \end{pmatrix} \quad (4)$$

According to the procedure above (the purpose is the method that is described in (Shi et al., 2011)), the local von-Neumann measurements on both qubits A and B are given by $\Pi_\pm^A = \frac{1}{2}(I \pm \vec{k}.\vec{\sigma})$ and $\Pi_\pm^B = \frac{1}{2}(I \pm \vec{l}.\vec{\sigma})$, where $\vec{k}$ and $\vec{l}$ are unit vectors in three-dimensonal real space. If $K = \vec{k}^T\vec{k}$ and $L = \vec{l}^T\vec{l}$, we can obtain MICC by using $\chi = \sum_{i,j=-,+}(\Pi_i^A \otimes \Pi_j^B)\rho(\Pi_i^A \otimes \Pi_j^B)$ for each of matrices, and (5) can be written for the square of HS (Shi et al., 2011):

$$D^2(\rho,\tau) = \frac{1}{4}\left[Tr(X+Y+TT^T) - Tr(XK+YL+TLT^TK)\right] \quad (5)$$

where (5) represents the square of distance between $R = \begin{pmatrix} 1 & \vec{y} \\ \vec{x}^T & T \end{pmatrix}$ and $R_\chi = \begin{pmatrix} 1 & \vec{y}L \\ K\vec{x} & KTL \end{pmatrix}$ ( for $\rho^{AB}$, $\vec{x}$ and $\vec{y}$ are the Bloch vectors of subsystem A and B respectively, and T is a $3 \times 3$ matrix and called correlation matrix (Shi et al., 2011)).

However, according to the general form of G, it can be argued that if the terms in equations (6) are satisfied, G for X and CS modes will have the same behavior:

$$\begin{cases} |p_2| = \left|\dfrac{2(q_1+q_3)-1}{4}\right| \\ |p_4| = \left|\dfrac{2(q_1+q_2)-1}{4}\right| \end{cases}, \quad p_7 = q_4, \quad p_6 = q_6, \quad q_2 + q_3 = \dfrac{1-\sqrt{16(p_1^2+p_3^2+p_5^2)-8p_1+1}}{2}$$

(6)

The equations that be mentioned above are obtained by the following way:

$$R_{\rho_X} = R_{\rho_{CS}}$$

In the calculations of G, $\lambda_M^{\max}$ (according to [13] quantity of the geometric measure is determined by $\lambda_M^{\max}$ or similar done by $\lambda_M^{\max}$) has particular importance, because of the mathematical, it cases abnormal behaviors of G. Thus, in two general forms it is determined as follows that just in original method (Shi *et al.*, 2011) some substitutes and placements are done;

*First case:* $|p_3| = |p_5| = 0$

As indicated in (Shi *et al.*, 2011), the most important quantity in $\lambda_M^{\max}$ set is $l'^2$ that for (3) with terms and conditions (6), is given by

$$l'^2 = 4p_6^2 - \left((2p_6)^2 - (4p_1-1)^2\right)l_3^2$$

If at all stages of selection $\lambda_M^{\max}$, $l_3$ and $l_1$ substitute, we can easily determine $\lambda_M^{max}$. In these calculations, if $p_7 = 0$, then $t_1 = t_2$, and if $p_7 \neq 0$, it will be $t_1 \neq t_2$.

*Second case:* $|p_3| \neq |p_5| \neq 0$

Here also replacing $l_3$ with $l_1$ is important.

$$l'^2 = 4(p_6+p_7)^2 l_1^2 + 4\left((p_6-p_7)-2p_3\right)^2 l_2^2 + (4p_1-4p_5-1)^2 l_3^2$$

In these calculations, if $p_7 = -p_3$, then $t_1 = t_2$, and if $p_7 \neq -p_3$, so it will be $t_1 \neq t_2$. Again we remind that all calculations related to the measure are presented in (Shi *et al.*, 2011) in detail.

## Physical examples

To better illustrate the topics discussed, let us mention two physical examples. In both of these examples, we arrive the density matrix of CS form and by using the above conditions (6), we will found X-state of it. Finally, geometric measure is obtained and it is plotted. The first example is related to the spin pair in a nanopore and a system consisting of N spin-carrying atoms of gas with spin s=1/2 in a closed nanopore in a strong external magnetic field B (Fel'dman *et al.*, 2012; Fel'dman & Rudavets, 2004), where the density matrix is obtained by the Hamiltonian of the averaged dipole-dipole interactions (Fel'dman & Rudavets, 2004)

$$H_{dz} = \frac{D}{2}\left(3I_z^2 - I^2\right)$$

where D is the coupling constant and is equal for all spin pair (Fel'dman & Rudavets, 2004; Baugh *et al.*, 2001), $I^2$ is the square of the total angular momentum, and $I_z = \sum_{i=1}^{N} I_i^z$ is the operator of the momentum projection of spin i (i= 1,2,3,…,N) on the axis Z (Fel'dman *et al.*, 2012).

If $\beta = \frac{\hbar \omega_0}{k_B T}$ ($\omega_0$ is the Larmour frequency, T is the temperature, and $k_B$ is the Boltzmann's constant), so $\rho_0 = \frac{1}{Z} e^{\beta I_x}$, where Z is the partition function, that $\rho_0$ is the density matrix in a strong external magnetic field. However, the time evolution of the density matrix is given by

$$\rho(t) = \frac{1}{Z} e^{-iatI_z^2} e^{\beta I_x} e^{iatI_z^2} \qquad (7)$$

where $a = \frac{3D}{2}$, and (7) describes the free induction decay and NMR line shape (Fel'dman & Rudavets, 2004; Fedorova *et al.*, 2009) for our purpose it shall be rewritten according to (Fel'dman *et al.*, 2012; Doronin *et al.*, 2007).

Accordingly, the reduced density matrix of the spin pair in a nanopore can be written as

$$\rho^{(1,2)}(t) = \begin{pmatrix} \frac{1}{4} & \frac{p}{2}-iu & \frac{p}{2}-iu & q-r \\ \frac{p}{2}+iu & \frac{1}{4} & q+r & \frac{p}{2}+iu \\ \frac{p}{2}+iu & q+r & \frac{1}{4} & \frac{p}{2}+iu \\ q-r & \frac{p}{2}-iu & \frac{p}{2}-iu & \frac{1}{4} \end{pmatrix}$$

where the correlation function as follows (Fel'dman *et al.*, 2012):

$$p = \frac{1}{2}\tanh\frac{\beta}{2}\cos^{N-1}(at),$$

$$\begin{cases} q+r = \dfrac{1}{4}\tanh^2\dfrac{\beta}{2} \\ q-r = \dfrac{1}{4}\tanh^2\dfrac{\beta}{2}\cos^{N-2}(2at) \end{cases},$$

$$u = \dfrac{1}{4}\tanh\dfrac{\beta}{2}\cos^{N-2}(at)\sin(at).$$

As regards, the density matrix is CS-type, so we can use the conditions (6) and find G that we have found for two modes with the difference coupling constant and the same number of spin N=100 (FIG.1 and FIG.2)

The second physical example related to spin-chain systems with Dzyaloshinskii-Moriya (DM) interaction (Dzyaloshinsky, 1958; Moriya, 1960) which the origin of it, is spin-orbit interaction (Aristov & Maleyev, 2000; Kargarian *et al.*, 2009; Jafari *et al.*, 2008). In this case, two-qubit anisotropic Heisenberg XXZ chain is considered with DM interaction parameter $D_x$ that its Hamiltonian is as follows (Chen & Zhi, 1998):

$$H = J\sigma_1^x \sigma_2^x + J\sigma_1^y \sigma_2^y + J_z \sigma_1^z \sigma_2^z + D_x \left( \sigma_1^y \sigma_2^z - \sigma_1^z \sigma_2^y \right) \tag{8}$$

where $D_x$ is the x-component parameter of the DM interaction, $J$ and $J_z$ are the real coupling coefficients, and $\sigma^i$ (i= x, y, z) are Pauli matrices.

As previously mentioned, the density matrix will be achieved by (8) the difference here is that changes are depending on temperature ($\rho(T) = \frac{1}{Z}\exp(-\beta H)$). $\rho(T)$ is the thermal state, where $Z = \text{tr}\left[\exp(-\beta H)\right]$ is the partition function of the system, and $\beta = \frac{1}{k_B T}$ with T temperature:

$$\rho'(T) = \frac{1}{2Z'} \begin{pmatrix} \mu_+ & -\xi & \xi & \mu_- \\ \xi & v_+ & v_- & -\xi \\ -\xi & v_- & v_+ & \xi \\ \mu_- & \xi & -\xi & \mu_+ \end{pmatrix}$$

where

$$\omega' = \sqrt{(J+J_z)^2 + 4D_x^2},$$

$$\phi = \arctan\left(\frac{2D_x}{J+J_z-\omega'}\right),$$

$$\varphi = \arctan\left(\frac{2D_x}{J+J_z+\omega'}\right),$$

and

$$\mu_\pm = e^{-\beta J_z} \pm \left(e^{\beta(J-\omega')} \sin^2\phi + e^{\beta(J+\omega')} \sin^2\varphi\right),$$

$$v_\pm = e^{-\beta(J_z - 2J)} \pm \left(e^{\beta(J-\omega')} \cos^2\phi + e^{\beta(J+\omega')} \cos^2\varphi\right),$$

$$\xi = i e^{\beta(J-\omega')} \sin\phi \cos\phi + i e^{\beta(J+\omega')} \sin\varphi \cos\varphi,$$

$$Z' = 2e^{-\beta J} \cosh\left[\beta(J-J_z)\right] + 2e^{\beta J} \cosh\left[\beta\omega'\right].$$

However, we can plot the geometric measure for the density matrix that is obtained (9). Graphs are plotted for different data $J$, $J_z$, and $D_x$ (FIG 3, FIG 4, and FIG 5).

# REFERENCES

**T. Yu and J.H.Eberly**. **2004.** A phenomenon termed entanglement sudden death (ESD), phys. Rev .Lett, B **93** ,140404.

**T. Yu and J.H.Eberly. 2006.** Sudden death of entanglement: classical noise effects, Opt. commun, 264, 393.

**T. Yu .2007.** Entanglement Decay Versus Energy Change: A Model, Phys. Lett. A **361**, 287,

**B. Bellomo, R. Lo Franco, and G. Compagno . 2007.** Non-Markovian effects on the dynamics of entanglement, Phys. Rev. Lett. **99**, 160502

**C.-L. Luo, L. Miao, X.-L. Zheng, Z.-H. Chen, and C.-G. Liao. 2011.** Sudden death of entanglement of two atoms interacting with thermal fields**,** Chin. Phys. B **20**, 080303.

**J. Ma, Z. Sun, X. Wang, and F. Nori, 2012**. Entanglement Dynamics of Two Qubits in a Common Bath , Phys. Rev. A **85**, 062323.

**M. Yönać, T. Yu, and J. H. Eberly, 2007.** Pairwise Concurrence Sudden Death: A Four-Qubit Model J. Phys. B **40**, S45.

**L. A Silva , et al. 2013.** Measuring bipartite quantum correlations of an unknown state**,** Phys. Rev. Lett. **110**, 140501.

**Cheng-Jie Zhang, et al. 2010.** Negative entanglement measure for bipartite separable mixed states, Phys. Rev. A **82**, 062312.

**H. Ollivier and W. H. Zurek. 2001.** Quantum Discord: A Measure of the Quantumness of Correlations, Phys. Rev. Lett. **88**, 017901.

**S. Luo. 2008.** Quantum discord of two-qubit X-states, Phys. Rev. A **77**, 042303.

**Mazhar Ali, A. R. P. Rau, G. Alber. 2010**. Quantum discord for two-qubit X states, Phys, Rev. A **81**. 042105.

**B. Dakic, V. Vedral, and C. Brukner. 2010.** Necessary and Sufficient Condition for Nonzero Quantum Discord, Phys. Rev. Lett. **105**, 190502 .

**S. Luo and S. Fu. 2010**. Geometric measure of quantum discord, Phys. Rev. A **82**, 034302.

**Mingjun Shi, Fengjian Jiang, Jiangfeng Du. 2011.** Symmetric geometric measure and dynamics of quantum discord, arXiv: quant-ph/1107.2958.

**M. A Yurischev. 2013.** Quantum discord for two-qubit CS states: Analytical solution, arXiv:1302.5239.

**Y.-X. Chen and Y. Zhi**. **1998**. Commun. Theor. Phys. 54, 02536102 .

**E. B. Fel'dman, E. I. Kuznetsova, and M. A. Yurischev 2012.** Quantum correlations in a system of nuclear s=1/2 spins in a strong magnetic field**,** J. Phys. A **45**, 475304**.**

**E.B. Fel'dman, M.G. Rudavets. 2004.** Nonergodic nuclear depolarization in nanocavities, J. Exp. Theor. Phys. **98**, 207.

**J. Baugh, A.Kleinhammes, D. Han, Q. Wang, and Y. Wu. 2001.** Confinement Effect on Dipole-Dipole Interactions in Nanofluids, Science **294**, 1505.

**A.V. Fedorova, E.B. Fel'dman, and D.E. Polianczyk. 2009**. Appl. Magn. Reson. 35, 511.

**S.I. Doronin, A.N. Pyrkov, E.B. Fel'dman. 2007**. Entanglement in alternating open chains of nuclear spins *s* = 1/2 with the *XY* Hamiltonian, JETP Letters **85**, 519.

**I. Dzyaloshinsky. 1958.** A thermodynamic theory of "weak" ferromagnetism of antiferromagnetics, J. Phys. Chem. Solids **4**, 241.

**T. Moriya. 1960.** Anisotropic Superexchange Interaction and Weak Ferromagnetism, Phys. Rev. **120**, 91.

**D.N. Aristov and S.V. Maleyev. 2000.** Spin chirality induced by the Dzyaloshinskii-Moriya interaction and polarized neutron scattering**,** Phys. Rev. B **62**, 751.

**M. Kargarian, et al. 2009.** Dzyaloshinskii-Moriya interaction and anisotropy effects on the entanglement of the Heisenberg model, Phys. Rev. A **79**, 042319.

**R. Jafari, et al. 2008.** Phase diagram and entanglement of the Ising model with Dzyaloshinskii-Moriya interaction, Phys. Rev. B **78**, 214414.

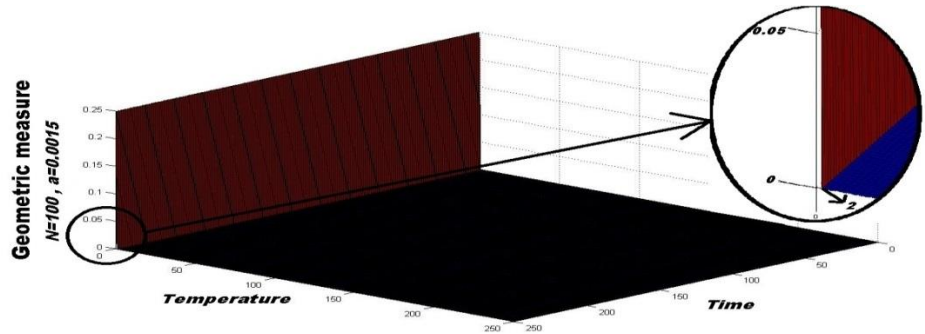

Figure 1: Schema of changes of G according to changes of time and temperature for number spin $N=100$ with $D=0.001$ (weak interactions)

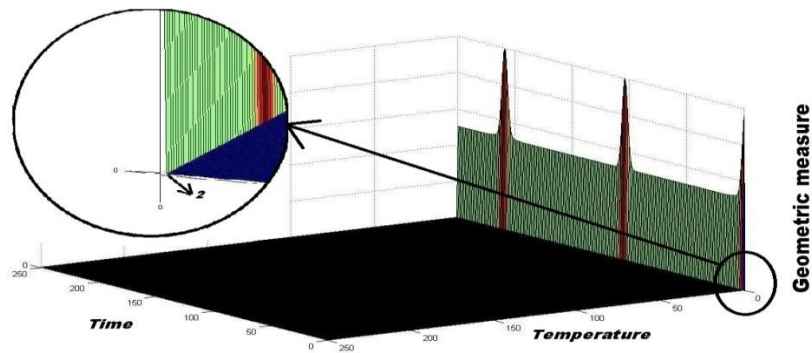

Figure 2: Schema of changes of G according to changes of time and temperature for number spin $N=100$ with $D=1$ (strong interactions)

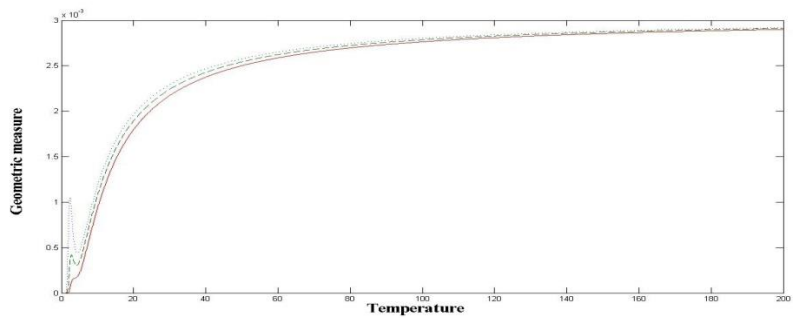

Figure 3: Schema of G according to changes of T with J=1, $D_x$=1 that both of them are constant, and $J_z$=0(china pointed line), $J_z$=0.4(dashed line), and $J_z$=0.9(normal line)

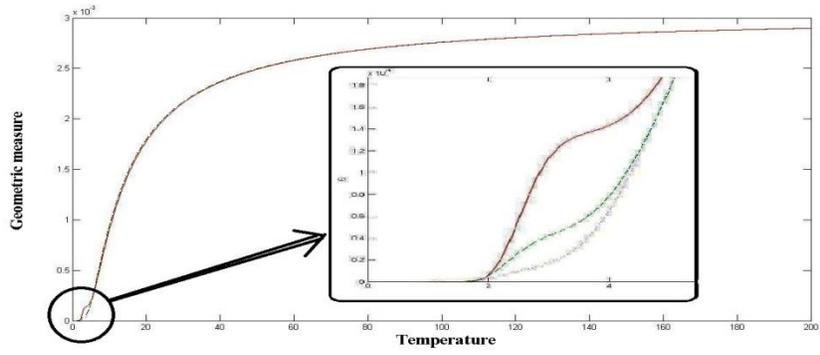

**Figure 4:** Schema of G according to changes of T with J=1, $J_z$=1 that both of them are constant, and $D_x$=0.5(china pointed line), $D_x$=0.7(dashed line), and $D_x$=1(normal line)

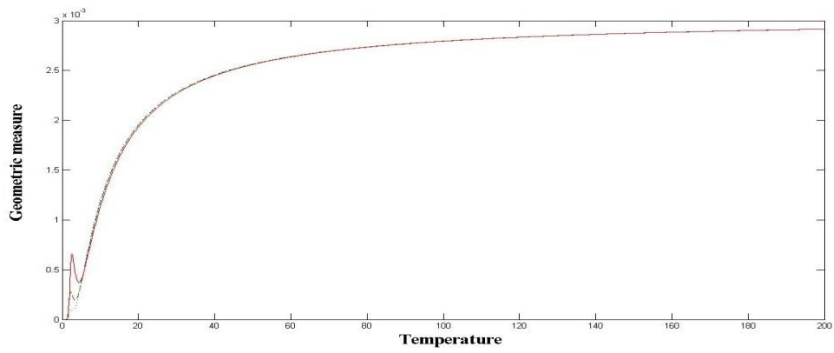

**Figure 5:** Schema of G according to changes of T with J=1, $J_z$=0.2 that both of them are constant, and $D_x$=0.5(china pointed line), $D_x$=0.7(dashed line), and $D_x$=1(normal line)